# First principles study on small $ZrAl_n$ and $HfAl_n$ clusters: structural, stability, electronic states and $CO_2$ adsorption

Hardik L. Kagdada[1], Shweta D. Dabhi[2], Venu Mankad[2], Satyam M. Shinde[3], Prafulla K. Jha[1*]

1. Department of Physics, Faculty of science, The M. S. University of baroda. Vadodara – 390002, INDIA
2. Department of Physics, M. K. Bhavnagar University, Bhavnagar – 364001, INDIA
3. School of Technology, Pandit Deendayal Petroleum University, Gandhinagar – 382007, INDIA

*prafullaj@yahoo.com

**Abstract**

We report a first principles study based on density functional theory on the structural and electronic properties of transition metal Zr and Hf doped small aluminum clusters with 1 to 7 aluminum atoms. We have used B3PW91/LANL2DZ basis set in Gaussian 09 package. The stability analysis reveals that the $ZrAl_4$ and $HfAl_4$ structures with $C_2v$ symmetry and square pyramid geometry are lowest energy structures. The most stable structures in $ZrAl_5$ and $HfAl_5$ are distorted tetrahedron type structure with symmetry $C_1$. The binding energies per atom for transition metal doped $Al_n$ clusters increases with the cluster size, while the second order difference in total energy show oscillatory behavior with even and odd cluster size. The HOMO – LUMO gap for $ZrAl_n$ is larger than the $HfAl_n$ clusters except for n = 1 and 3. The $HfAl_6$ has more tendency to accept or give away electrons. The negative charge exists on Zr and Hf indicating that the electron transfers from Al atom to transition metal, Zr and Hf. The thermodynamical analysis suggest that the $HfAl_6$ cluster has highest exothermicity compared to not only all considered Al clusters but also other transition metal doped Al clusters reported in J. Phys. Chem. C, 120, 10027 (2016).

## 1. Introduction

In recent years great deal of attention is devoted to clusters due to their fundamental concept of electronic and geometrical stability and potential application in variety of fields such as catalytic activities, hydrogen storage[1 – 4], formation of metallic glasses[5, 6] and most importantly alternate building blocks of materials in the form of super atoms[7]. Among the variety of clusters, binary metallic clusters have much attention due to their wide range of tailoring properties about its size, shape and chemical composition[8 – 12]. Furthermore, binary metallic clusters are used to control the synthesis and chirality of the single walled carbon nano tubes (SWCNT)[13]. As a typical example of atomic and doped atomic clusters, the aluminum and transition metal (TM) doped aluminum clusters respectively have been the subject of many investigations[14 – 29]. The display of large local magnetic moments by the transition metal doped aluminum clusters is one of the main reason for their increasingly focused studies[29]. The existence of large local magnetic moments is attributed to the interaction of d electron with the nearly free electron gas when the TM element is present in sp metal host or surfaces. Wang et. Al[29] studied the structural, electronic and magnetic properties of $MAl_n$ (M = Cr, Mn, Fe, Co, Ni; n = 1 – 7, 12) clusters using density functional theory (DFT) by treating the exchange-correlation interaction with generalized gradient approximation (GGA). While they found non-degenerate and delocalized HOMO and LUMO states, the computed total magnetic moment oscillate with the cluster size. Very recently, the influence of spin on the properties of small sized transition metal doped aluminum clusters using DFT is reported[7]. Their low spin doped aluminum clusters show the odd even oscillation in various calculated properties in the line of jellium shell structures in contrast to the high spin case which show smooth variation in the properties. However in comparison, Zr and Hf based transition metal doped Al clusters have not been studied systematically and lacks the information on stability, structures, electronic, heat of absorption on the cluster surface. Zr and Hf transition metal elements possess many interesting properties like high coercivity and low absorption cross section for neutrons.

There exist few studies based on Zr-Al compounds using DFT calculations[30 – 35]. Wang et. al[30]. studied the phase stability, mechanical and thermodynamical properties of Zr-Al binary substitutional alloys and found that the $ZrAl_2$ is most stable with excellent mechanical properties, while $ZrAl_3$ possesses better thermal conductivity and higher melting temperature[30]. Arikan[31] has reported the elastic, electronic and phonon properties of $Zr_3Al$ compound. A stability and thermodynamic property of $Zr_2Al$ under high pressure is reported using DFT method within GGA for exchange and correlation[35]. De Souza[32] et. al. performed a basin-hopping Monte Carlo investigation within the embedded-atom method for the structural and energetic properties of bimetallic ZrCu and ZrAl nanoclusters with 55 and 561 atoms and found that unary systems adopt the well-known compact icosahedron (ICO) structure. However, the ICO structure changes to the nearly spherical shape due to strong minimization of the surface energy when both chemical species change towards a more balanced compositions. DFT study of structural and electronic properties of $Zr_nAl^{\pm m}$ (n = 1 – 7, m = 0, 1) clusters shows that all stable structures are three dimensional for n > 3 and binding increases as n increases for all considered $Zr_nAl^{\pm m}$ [33].

In recent time the sequestration of the $CO_2$ emitted from industrial manufacturing plants is one of the most pressing issues in the environmental protection. An ideal $CO_2$ sequestration material should have large surface areas and strong adsorption sites. Many $CO_2$ adsorbents including metal organic frameworks (MOFs), carbon and silicon carbide and boron rich boron nitride have limitations in terms of interactions[36 – 40]. Clusters have been found good not only to absorb the $CO_2$ but also $H_2$, $O_2$, and $N_2$. Sengupta et. al. [7] have studied the $CO_2$ absorption over transition metal series from Sc to Zn doped aluminum clusters ($Al_5$ and $Al_7$) and found that the starting element of series Sc and Ti doped Al clusters are good absorber. Therefore, it would be of interest to see the absorption capability of non-magnetic and closer to Sc and Ti elements, Zr and Hf doped aluminum clusters.

In the present paper, on the basis of results of first principles calculations using theories of plane waves and localized atomic orbitals we report relative stability, binding energy and electronic properties of Zr and Hf doped Al clusters. Furthermore, we report that the Zr and Hf doped Al clusters can capture $CO_2$ more strongly than the reported other transition metal doped Al clusters. The structures of optimized clusters and $CO_2$ clusters conformers are found to be consistent with available similar $CO_2$ cluster conformers.

## 2. Computational method

In the present study all the calculations were performed using density functional theory simulation on $TmAl_n$ clusters using GAUSSIAN 09 package[41]. The Becke's three parameter (B3) and Perdew and Wang (PW91) GGA functional together called the B3PW91 were used for the exchange and correlation respectively along with Los Alamos set of double-zeta type (LANL2DZ)[42 – 44] basis set to derive a complete geometrical optimization of transition metal Zr and Hf doped $Al_n$ (n = 1 – 7) clusters. The accurate determination of ground state geometry of cluster is required in the chemistry of clusters. Initially different arrangements of isomers of clusters are taken for each cluster to obtain the ground state conformers. After optimization of all these isomers the energetically minimum structure is taken as ground state geometry for each cluster. Binding energy between transition metal and $Al_n$ clusters is calculated using the following formula

$$E_b(n) = \frac{[nE(Al) + nE(Tm) - E(TmAl_n)]}{n+1} \quad (1)$$

where E(Al) and E(Tm) is the total energy of one atom of Al and transition metal (Zr, Hf) respectively and $E(TmAl_n)$ is the total energy of transition metal doped $Al_n$ cluster. The relative stability of the cluster is defined by the second order difference in the total energy which was calculated by the following formula

$$\Delta^2 E = E(TmAl_{n-1}) + E(TmAl_{n+1}) - 2E(TmAl_n) \quad (2)$$

Chemical hardness defines the tendency to gain or give the electrons is calculated using the following formula[45, 46]

$$\eta = \frac{I.E. - E.A.}{2} \quad (3)$$

where I.E. = Ionization Energy and E.A. = Electron Affinity. The difference in energy of neutral cluster and cation of that cluster are taken to evaluate the ionization energy and the difference in energy of neutral cluster and anion of that cluster are taken to evaluate the electron affinity. The behavior of charge transfer between transition metal and Al cluster, natural electronic configuration are calculated using natural bond analysis (NBO) which is implemented in Gaussian 09 package[47, 48].

## 3. Result and Discussion

### *3.1 Structural Geometry:*

For calculation of any ground state properties one needs to optimize the structures and obtain the equilibrium geometrical structures. Figs. 1 and 2 present the fully optimized ground state and low lying structures of transition metal doped $Al_n$ clusters ($TmAl_n$; TM = Zr and Hf, n = 1 to 7). The point group and the multiplicity are listed in Table – I. The Table – I also lists the binding energy, second order difference in total energy and ionization energy for $ZrAl_n$ and $HfAl_n$ clusters. Relatively lowest energy structure is found for n = 2 i.e. for $ZrAl_2$ and $HfAl_2$ with the point group symmetry Cs and $C_2v$ for $ZrAl_2$ and $HfAl_2$ respectively. They have a shape of isosceles triangle with angle $58.164^o$ for Al-Zr-Al and $59.585^o$ for Al-Hf-Al. For n = 3 minimum energy structure is found for Cs symmetry while other $ZrAl_3$ and $HfAl_3$ structures are 2.31 eV and 2.66 eV higher in energy, respectively. $ZrAl_4$ and $HfAl_4$ structures with $C_2v$ symmetry and square pyramid geometry are lowest energy structures while others have distorted tetrahedral $C_1$ symmetry with higher energy. The most stable structure in $ZrAl_5$ and $HfAl_5$ are distorted tetrahedron type structure with symmetry $C_1$ while other structures of $ZrAl_5$ and $HfAl_5$ with $C_2v$ symmetry are 1.74 eV and 1.85 eV higher in energy. $TmAl_6$ structures have both $C_4v$ and $C_2v$ type symmetry. The $C_4v$ type structures have 2.13 eV and 1.24 eV higher energy than that of $C_2v$ type structure for $ZrAl_6$ and $HfAl_6$ respectively. The structures with pentagonal bi pyramid $C_2v$ are ground state structures for $TmAl_6$ which is similar to the $FeAl_6$ clusters[29]. The ground state isomers of $ZrAl_7$ and $HfAl_7$ are to be found by capping of Al atom on the pentagonal bipyramid structures of $TmAl_6$ and having symmetry $C_2v$. All calculations are performed for the most energetically stable structures.

### *3.2 Stability*

For the analysis of relative stability of $TmAl_n$ clusters we have calculated binding energy per atom $E_b(n)$ and $2^{nd}$ order difference in the total energy $\Delta^2E$ of the clusters using the equations (1) and (2) respectively presented in section 2. The second order difference in the total energy is the quantity which defines the relative stability of the clusters[49]. The calculated values of $E_b(n)$ and $\Delta^2E$ for ground state structures are listed in Table – I. The binding energy per atom increases with the increasing in cluster size as can be seen from Fig. 3(a). Table – I and Fig. 3(b) show that the $\Delta^2E$ behaves oscillatory with cluster size. The fluctuation behavior of $\Delta^2E$ is more for $HfAl_n$ than $ZrAl_n$. However, we observe that the $\Delta^2E$ is higher for clusters of even number, which is similar to the clusters of $Zr_nAl^{\pm m}$ [33]. The maximum peak observed at n = 4 for $HfAl_4$ suggest that $HfAl_4$ is the most stable cluster than its nearest neighbors (n = 3, 5). This can also be seen from the Fig. 3(a) that the $HfAl_4$ has highest binding energy. As for n = 6, $ZrAl_n$ is concerned that the n = 6 is most stable, which can also be seen from the Fig. 3(a) where binding energy is highest for $ZrAl_6$. All the calculations presented in the following sections are performed for the most energetically stable structures.

*3.3 Electronic properties*

Electronic properties of clusters can be analyzed through Highest Occupied Molecular Orbital (HOMO) and Lowest Unoccupied Molecular Orbital (LUMO) orbitals. HOMO – LUMO orbital play an important role in the process of chemical reactions hence, its analysis can help in studying the chemical stability of the clusters. The molecule can be easily excited if the gap is smaller. Fig. 4(a) shows the HOMO – LUMO gap ($G_{HL}$) for $TmAl_n$ clusters as a function of cluster size n. The HOMO – LUMO gap for $ZrAl_n$ is larger than the $HfAl_n$ clusters except for n = 1 and 3. Maximum and minimum gap is found for HfAl and $HfAl_4$ respectively with the value of 2.214 eV and 0.159 eV which indicates that the HfAl cluster has relatively higher chemical stability and $HfAl_4$ is more inert cluster than others. Chemical hardness ($\eta$) as a function of cluster size (n) is presented in Fig. 4(b) and listed in Table – II. Chemical hardness is the measure of the cluster to accept or give away electrons i.e. bigger $\eta$ suggest the smaller tendency and smaller $\eta$ suggest the larger tendency for accepting or give away the electrons[50]. Table – II shows that the $HfAl_6$ has minimum $\eta$ with 1.825 eV indicating that the $HfAl_6$ has more tendency to accept or give away electrons.

*3.4 NBO analysis*

The natural electronic configuration and charge transfer between considered transition metals and Al cluster can be studied by natural bond analysis (NBO). The natural electronic configuration and atomic charge on each atom in $ZrAl_n$ and $HfAl_n$ cluster presented in supplementary Table – I. The valance electron orbit for Zr, Hf and Al is $4d^2\ 5s^2$, $5d^2\ 6s^2$ and $3s^2\ 3p^1$ respectively. The range of electrons in 3s orbital of Al in $ZrAl_n$ is from 1.28 to 1.84 and that in $HfAl_n$ within 1.21 to 1.81, which shows that the 3s orbital of Al losses the electrons. While for 3p orbit of Al gains the electrons as the occupation of electrons in 3p orbital of Al in $ZrAl_n$ within 1.07 to 1.97 and that for the $HfAl_n$ has range of 1.11 to 1.93. The contribution of 4p orbitals of Al in both type of clusters can be neglected as the electrons in this orbit is in the range of 0.01 – 0.02. This shows that the hybridization of Al in Zr and Hf doped $Al_n$ is sp hybridization. The electrons in 4d orbitals of Zr is between 2.61 and 4.14, while for electron in 5s orbital of Zr within 0.47 to 1.32. This indicates that the 5s orbital of $ZrAl_n$ clusters losses the electrons, while 4d orbitals occupied more electrons. The occupation nature of electrons is similar for $HfAl_n$ clusters as 5d orbital gain the electrons while 6s orbit losses the electrons. It is observed from supplementary Table – I that occupation of electrons in 5p orbital of Zr is in between 0.10 and 0.62, while for $HfAl_n$ within 0.21 to 0.75. This occupation shows that the Zr and Hf has spd hybridization in $Al_n$ clusters. The natural electronic configuration indicates that the loss of electrons in 3s orbitals of Al is more that the gain of electrons in 3p orbital. However, the loss of electrons in 5s orbitals of Zr is less than the occupied electrons in 4d orbital of Zr in $ZrAl_n$ clusters and similar phenomena happens in $HfAl_n$. This indicates that the charge transfers from aluminum to Zr and Hf respectively in $ZrAl_n$ and $HfAl_n$ clusters and both Zr and Hf have negative charge as can be seen from Table – II.

### *3.5 Adsorption of $CO_2$ on $TmAl_n$ Clusters*

We have computed the thermodynamic data of the adsorption of $CO_2$ on the Hf and Zr doped $Al_n$ (n = 4 – 7) clusters and compare them with the available data on other transition metal doped Al clusters[7]. The literature reveals that there are mainly two different configuration of bonding between $CO_2$ and $TmAl_n$ clusters. One is series bonding i.e. the one O atom from $CO_2$ bind with the transition metal of cluster and other is the parallel bonding i.e. the one oxygen atom bind with transition metal and other oxygen atom bind with the one aluminum cluster. In the present calculation we take parallel configuration because Sengupta[7] et. al. have reported that the series binding has low exothermicity of adsorption. Figs. 5 (a-h) show the optimized structures of transition metal Hf and Zr doped $Al_n$ clusters with adsorption of $CO_2$ in parallel binding configuration. Thermochemical parameters (ΔE, ΔH and ΔG) are listed in Table – III which depicts that the $HfAl_6$ cluster has highest exothermicity compared to the other Hf and Zr doped $Al_n$ clusters. It is interesting to note that the exothermicity for $HfAl_6$ is even higher than the Sc and Ti doped aluminum clusters[7]. Furthermore, one can observe that in all cases the angle of $CO_2$ is changed and the bond length between $CO_2$ and cluster is increased as observed in the case of other transition metal doped $Al_n$ clusters.

## 4. Conclusions

In summary we have performed the density functional theory calculations with the basis set of LANL2DZ of two transition metals (Zr and Hf) doped aluminum clusters. We report energetically stable geometries, electronic structure and relative stability of Zr/Hf $Al_n$ (n = 1 – 7) clusters. All the calculations are performed for the most energetically stable structures. The Zr and Hf doped $Al_n$ clusters favor 3D spatial structures at the smaller number of Al atoms compared to higher number of Al clusters. The natural electronic configuration analysis shows that the electron moves from $Al_n$ clusters to transition metal Zr and Hf. We found that the binding energy of considered clusters is increasing with increase in cluster size. $HfAl_4$ is relatively stable structure. From the HOMO – LUMO analysis we found that the ZrAl is relatively chemically more active with HOMO – LUMO gap of 2.124 eV. $HfAl_6$ cluster has more tendencies to accept or give away the electrons and higher enthalpy difference as well as higher adsorption of $CO_2$.

## 5. Supplementary Materials

See supplementary material for the natural bond analysis (NBO). The natural electronic configuration and atomic charge on each atom in $ZrAl_n$ and $HfAl_n$ cluster is presented in supplementary Table – I.

## Acknowledgements

Authors are thankful to the SERB and DST for providing computational facility through project and FIST programs respectively. SD thanks DST for INSPIRE fellowship while VM thanks SERB for young scientist award.

**Figure caption**

**Figure – 1:** Lowest energy and low lying energy structures of $ZrAl_n$, (n = 1 to 7). Here first digit denotes the number of aluminum and the characters (a, b, c) shows the structure label (lowest energy to highest energy structures.) Red color is for Zirconium and blue for the aluminum.

**Figure – 2:** Lowest energy and low lying energy structures of $HfAl_n$, n = 1 to 7. Here first digit denotes the number of aluminum and the characters (a, b, c) shows the structure label (lowest energy to higher energy structures) Cyan color is for hafnium and blue color for the aluminum).

**Figure – 3:** (a) Binding energy and (b) second order difference in total energy for $ZrAl_n$ and $HfAl_n$ clusters.

**Figure – 4:** (a) HOMO-LUMO gap and (b) Chemical hardness for $ZrAl_n$ and $HfAl_n$ clusters.

**Figure – 5:** Ground structures of $CO_2$ dopped $ZrAl_n$ (a, b, c, d), $HfAl_n$ (e, f, g, h), n = 4 – 7 and (i) $CO_2$ structure.

**Figure – 6:** Thermochemical data on absorption of $CO_2$ for (a) $ZrAl_n$ and (b) $HfAl_n$

**Table caption**

**Table – I:** Structural symmetry, binding energy and second order difference in total energy with ionization energy for the $ZrAl_n$ and $HfAl_n$ clusters.

**Table – II:** HOMO-LUMO gap, Atomic charge on Tm and chemical hardness for $ZrAl_n$ and $HfAl_n$.

**Table – III:** Thermochemical data (enthalpy and free energy difference) of adsorption of $CO_2$ on $ZrAl_n$ and $HfAl_n$ (n = 4 to 7).

.

**Figure – 1**

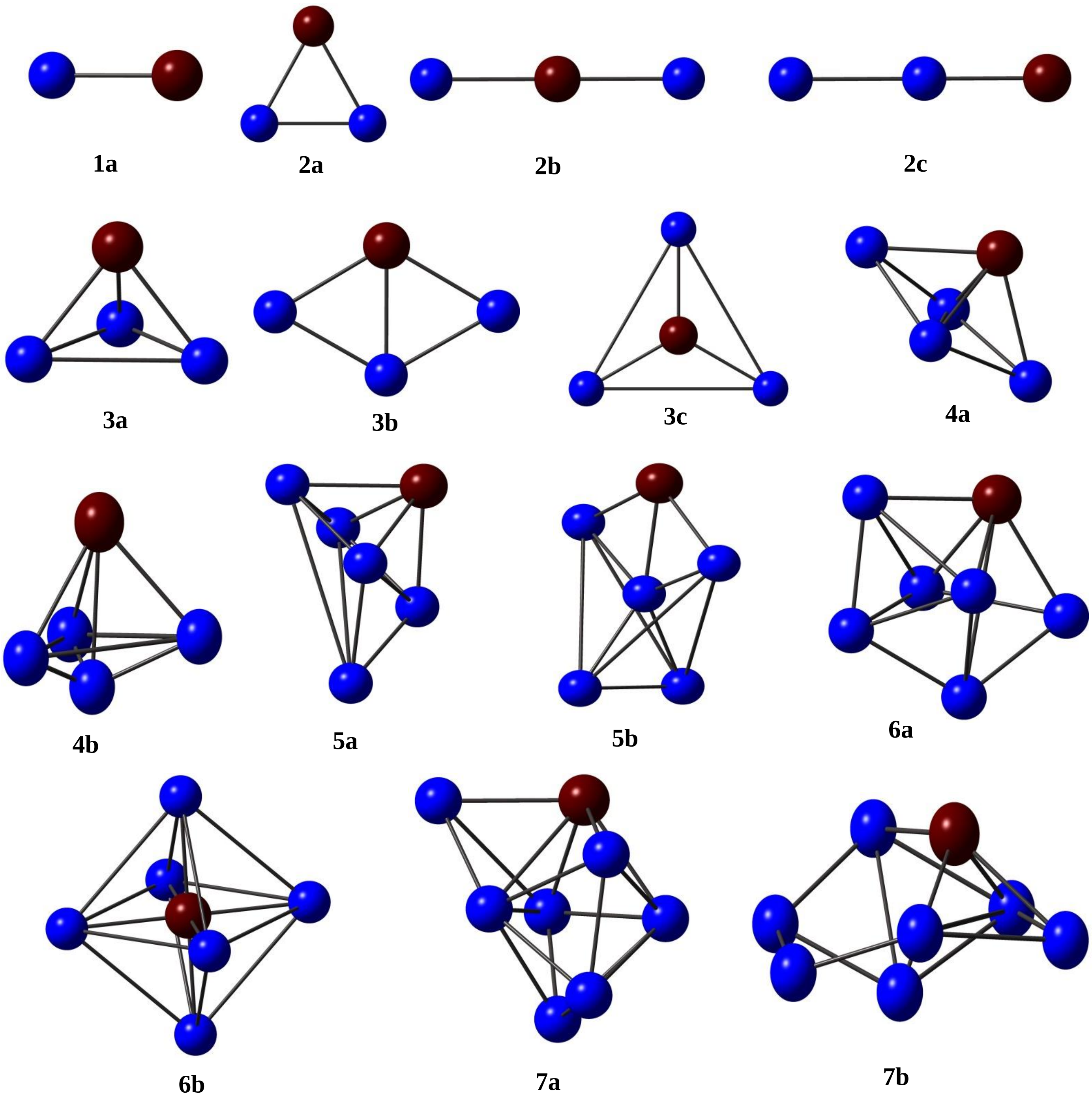

**Figure – 2**

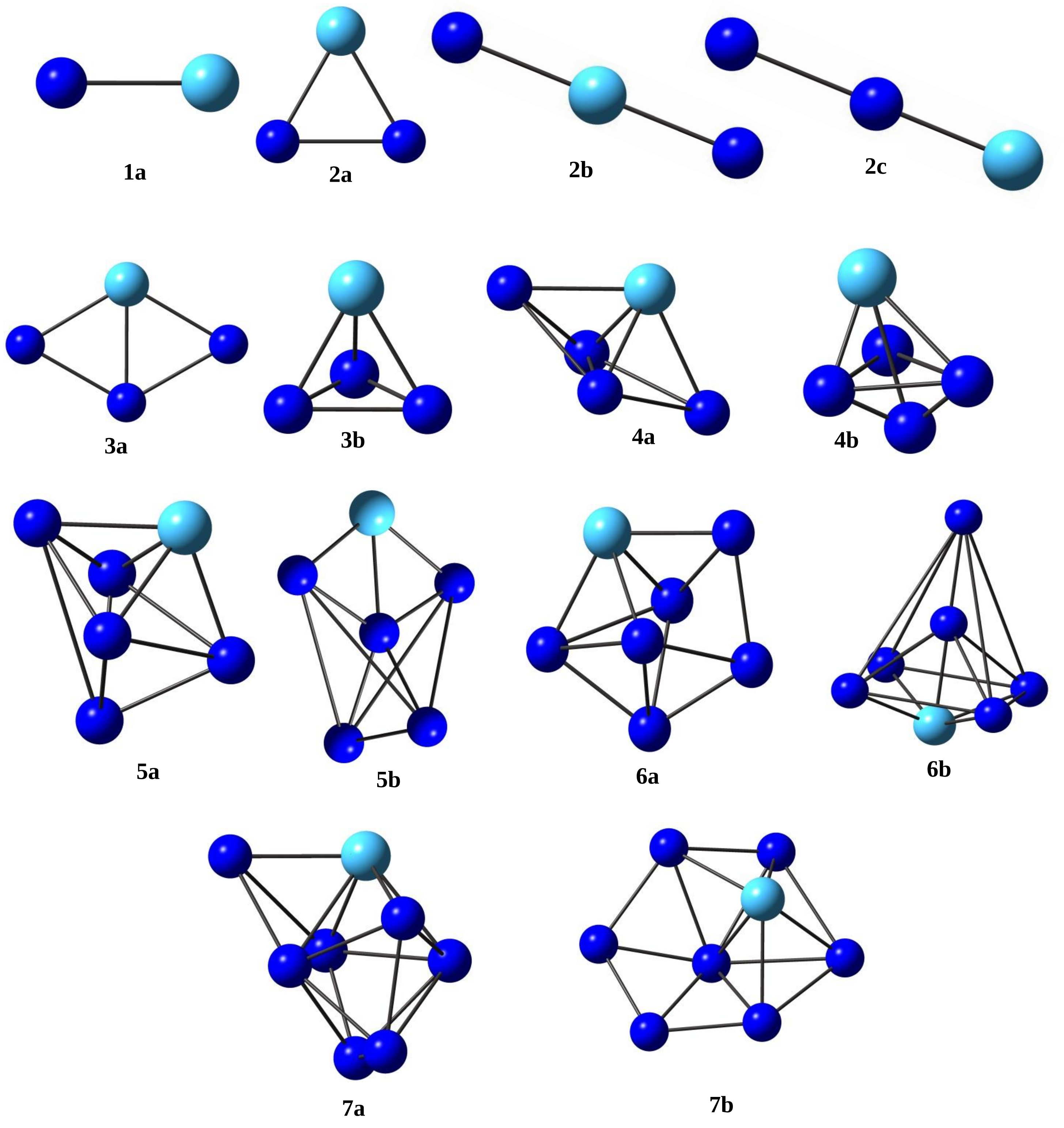

**Figure – 3**

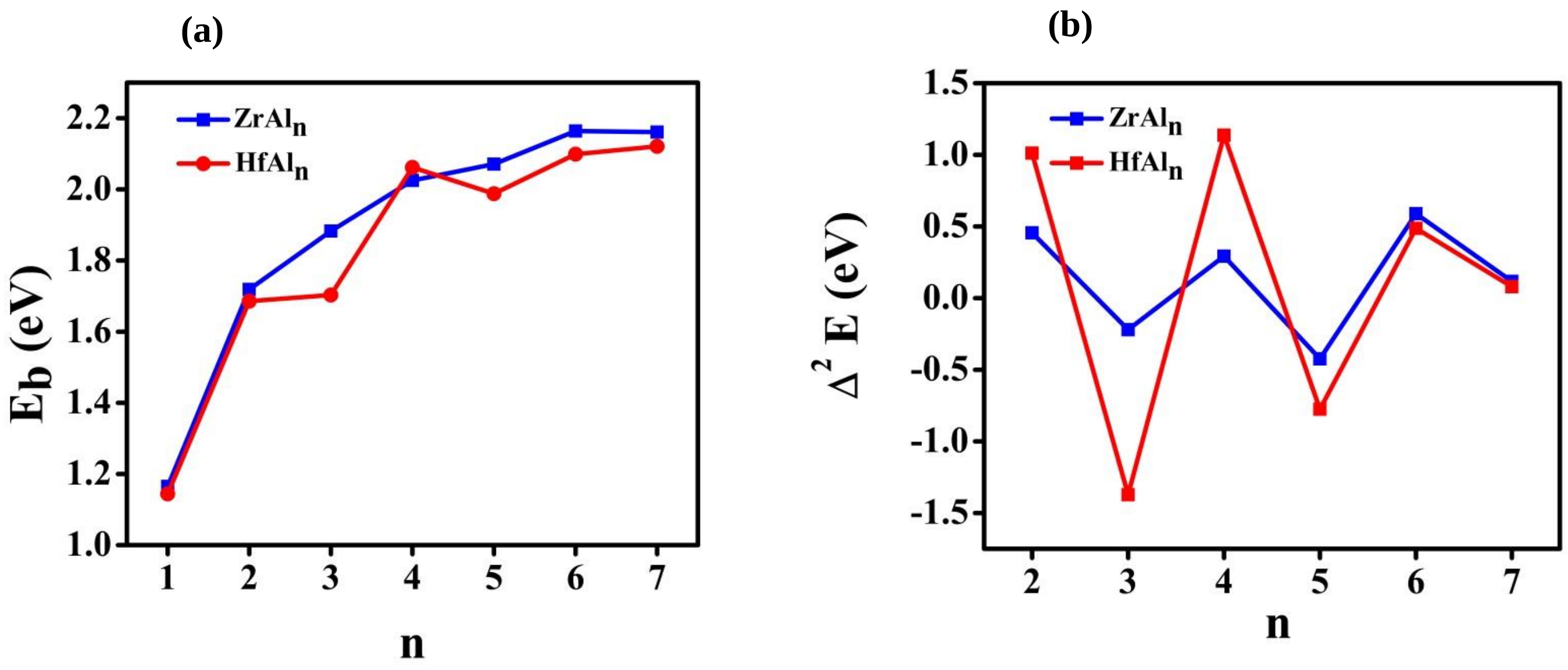

**Figure – 4**

(a)

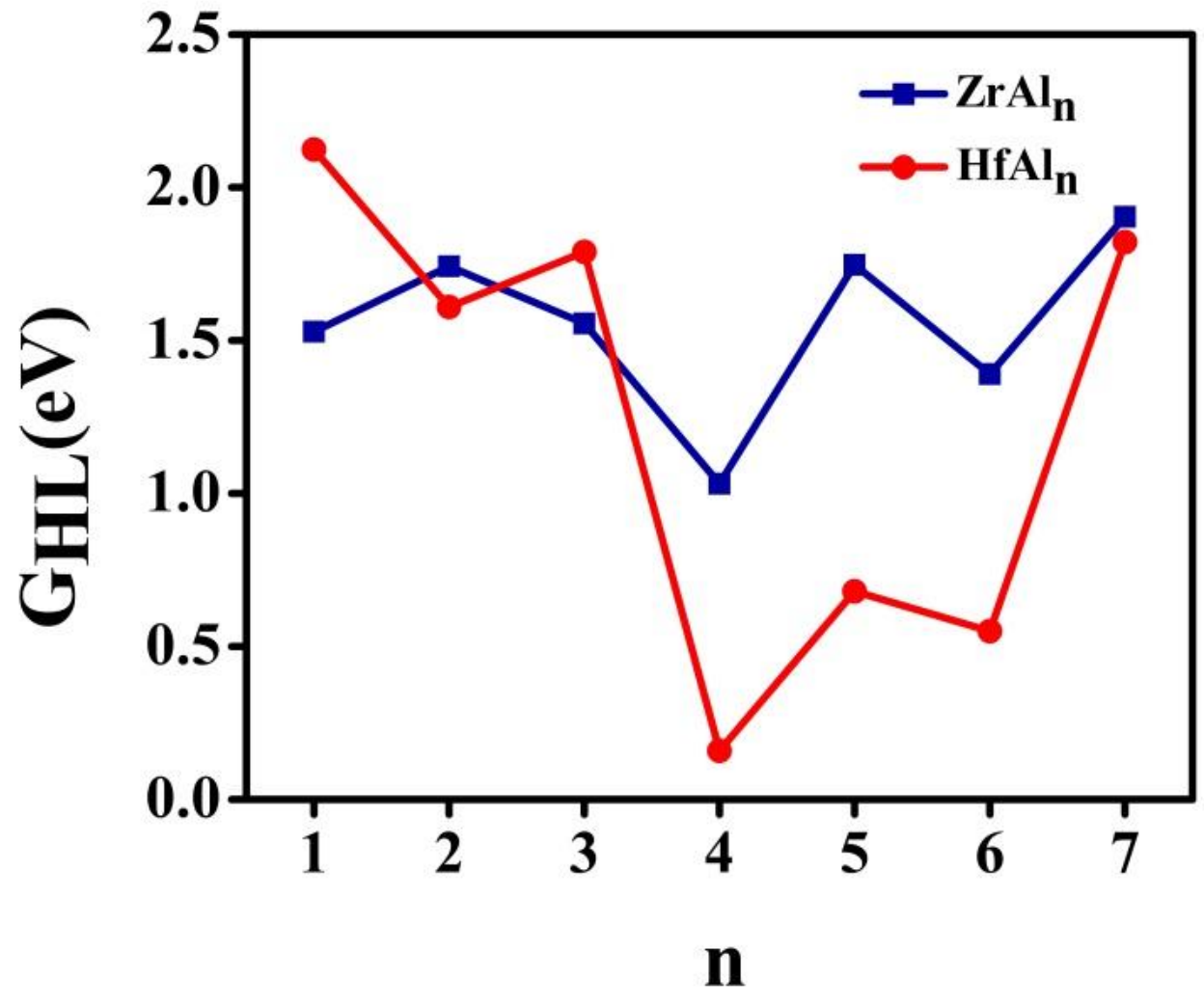

(b)

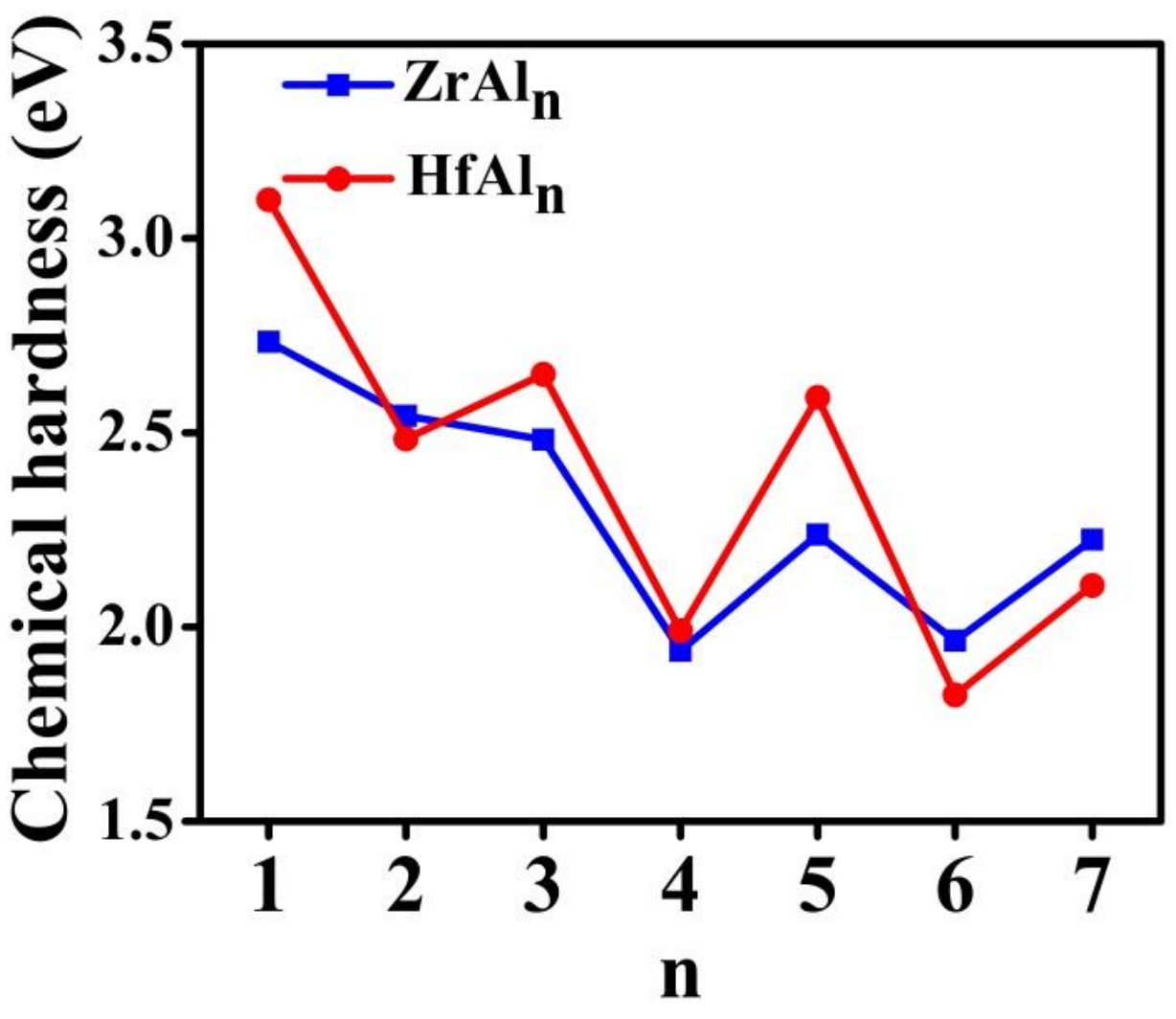

**Figure – 5**

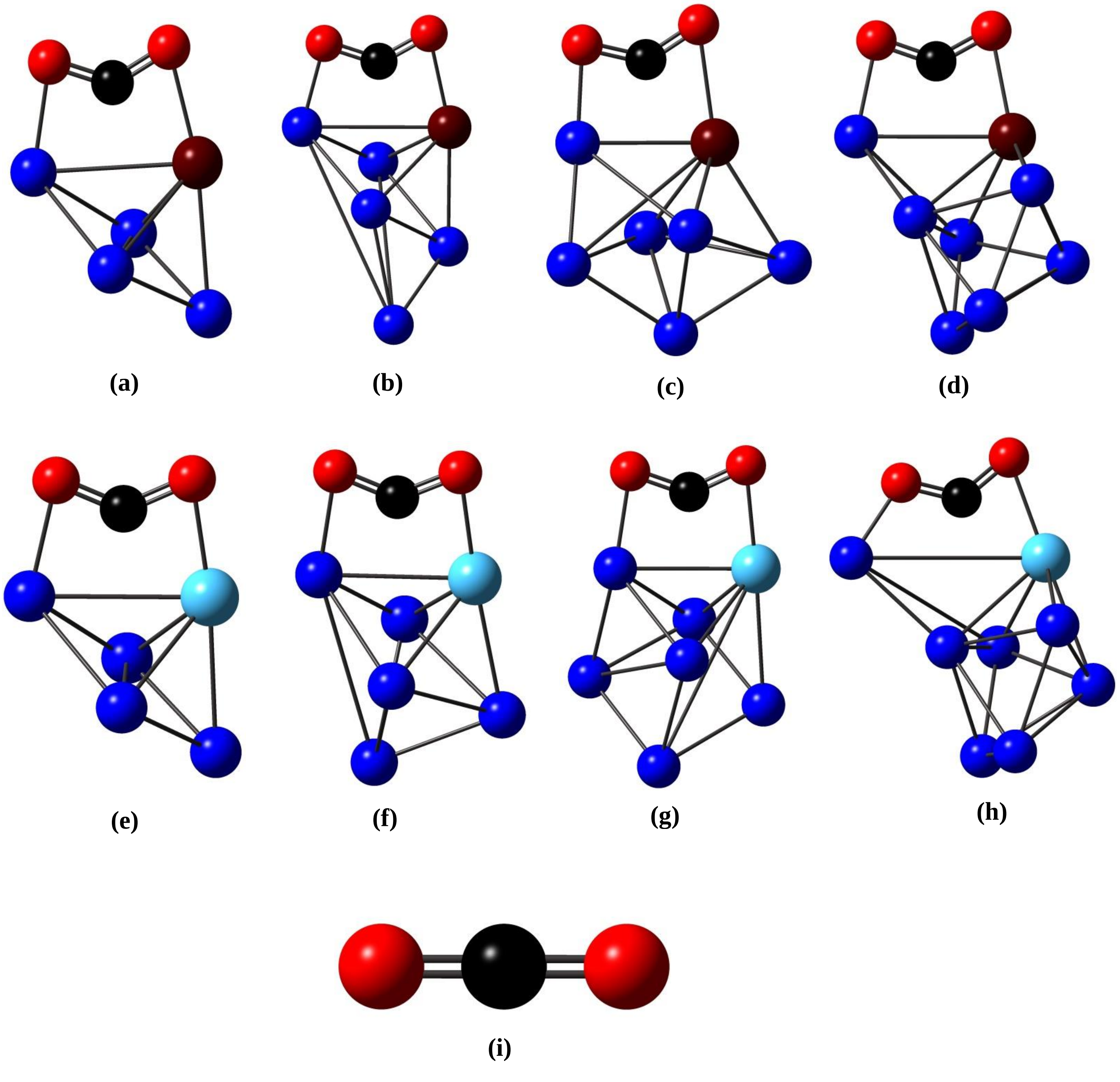

**Figure – 6**

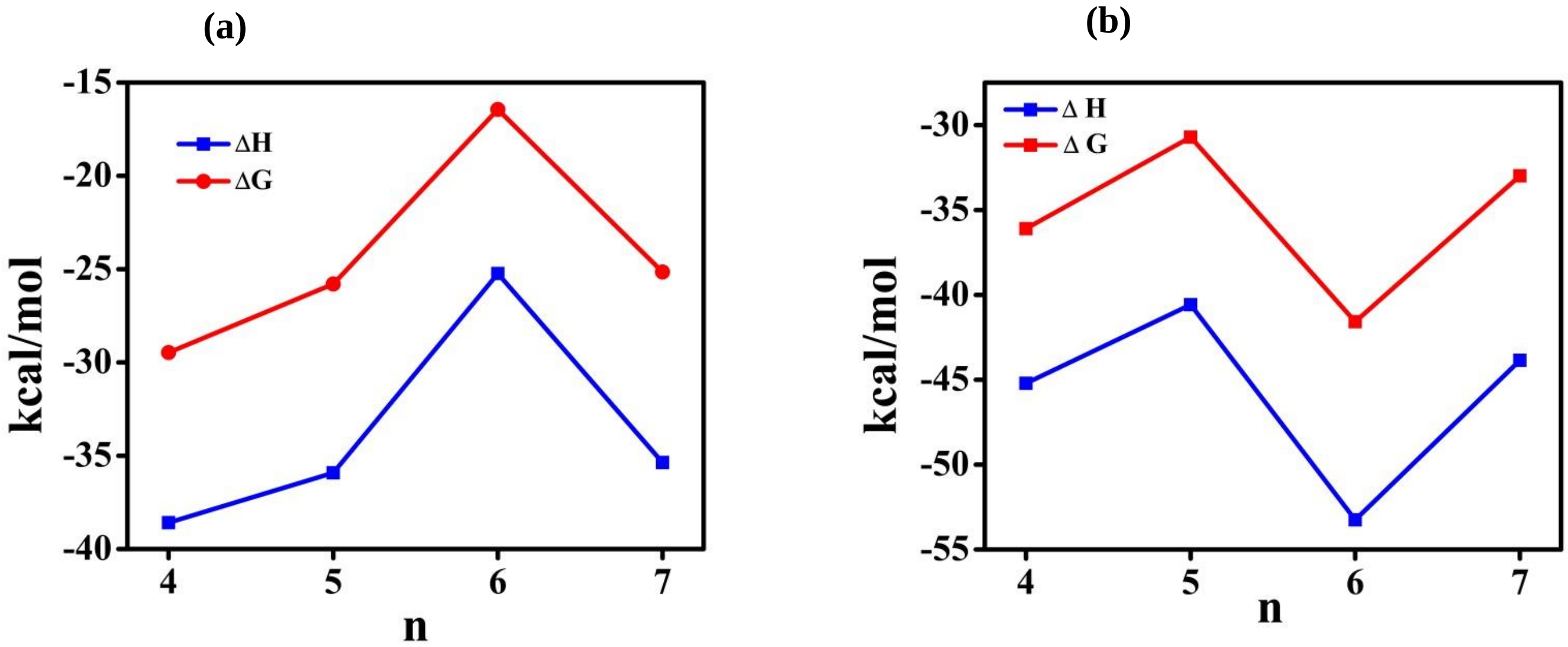

**Table – 1: Structural symmetry, binding energy and 2nd order difference in total energy with ionization energy for the $ZrAl_n$ and $HfAl_n$ clusters**

| Clusters(multiplicity) | Symmetry | $E_b$ (eV) | $\Delta^2E$ (eV) | Ionization Energy (eV) |
|---|---|---|---|---|
| ZrAl | $C_\infty v$ | 1.165 | --- | 6.565 |
| $ZrAl_2$ | $C_s$ | 1.719 | 0.456 | 6.293 |
| $ZrAl_3$ | $C_s$ | 1.883 | -0.220 | 6.501 |
| $ZrAl_4$ | $C_2v$ | 2.025 | 0.293 | 5.931 |
| $ZrAl_5$ | $C_1$ | 2.071 | -0.424 | 6.060 |
| $ZrAl_6$ | $C_2v$ | 2.164 | 0.589 | 6.049 |
| $ZrAl_7$ | $C_1$ | 2.161 | 0.117 | 6.494 |
| HfAl | $C_\infty v$ | 1.144 | --- | 7.157 |
| $HfAl_2$ | $C_2v$ | 1.686 | 1.012 | 6.402 |
| $HfAl_3$ | $C_s$ | 1.703 | -1.372 | 6.833 |
| $HfAl_4$ | $C_2v$ | 2.062 | 1.137 | 6.032 |
| $HfAl_5$ | $C_1$ | 1.988 | -0.776 | 6.789 |
| $HfAl_6$ | $C_2v$ | 2.099 | 0.487 | 6.003 |
| $HfAl_7$ | $C_1$ | 2.121 | 0.080 | 6.256 |

**Table – 2: HOMO-LUMO gap, Atomic charge on Tm and chemical hardness for $ZrAl_n$ and $HfAl_n$**

| Clusters | HOMO-LUMO gap (eV) | Atomic charge (a.u.) | Chemical hardness (eV) |
|---|---|---|---|
| ZrAl | 1.529 | -0.090 | 2.734 |
| $ZrAl_2$ | 1.742 | -0.128 | 2.543 |
| $ZrAl_3$ | 1.555 | -0.871 | 4.965 |
| $ZrAl_4$ | 1.032 | -1.061 | 1.939 |
| $ZrAl_5$ | 1.747 | -0.715 | 2.238 |
| $ZrAl_6$ | 1.391 | -1.122 | 1.965 |
| $ZrAl_7$ | 1.903 | -1.324 | 2.225 |
| HfAl | 2.214 | -0.073 | 3.099 |
| $HfAl_2$ | 1.609 | -0.182 | 2.484 |
| $HfAl_3$ | 1.789 | -0.945 | 2.650 |
| $HfAl_4$ | 0.159 | -1.345 | 1.990 |
| $HfAl_5$ | 0.680 | -1.095 | 2.591 |
| $HfAl_6$ | 0.549 | -1.002 | 1.825 |
| $HfAl_7$ | 1.822 | -1.223 | 2.107 |

**Table – 3: Thermochemical data (enthalpy and free energy difference) of adsorption of $CO_2$ on $ZrAl_n$ and $HfAl_n$ (n = 4 to 7).**

| Clusters | ΔE (kcal/mol) | ΔH (kcal/mol) | ΔG (kcal/mol) |
|---|---|---|---|
| $ZrAl_4$ | -39.152 | -38.587 | -29.465 |
| $ZrAl_5$ | -35.097 | -35.912 | -25.796 |
| $ZrAl_6$ | -25.065 | -25.232 | -16.443 |
| $ZrAl_7$ | -35.235 | -35.363 | -25.148 |
| $HfAl_4$ | -45.106 | -45.215 | -36.096 |
| $HfAl_5$ | -40.467 | -40.580 | -30.708 |
| $HfAl_6$ | -53.240 | -53.253 | -41.580 |
| $HfAl_7$ | -43.798 | -43.855 | -32.991 |